\documentclass[twocolumn]{aastex62}
\usepackage{amsmath}
\usepackage{CJK}

\newcommand{\dif}{{\rm d}}

\newcommand{\barnp}{{\bar{n}_{\rm p}}}
\newcommand{\Fp}{{F_{\rm p}}}

\shorttitle{Metallicity Effect on \emph{Kepler} Planets}
\shortauthors{Wei Zhu}

\begin{document}
\begin{CJK*}{UTF8}{gbsn}

\title{Influence of Stellar Metallicity on Occurrence Rates of Planets and Planetary Systems}

\author{Wei~Zhu (祝伟)}
\affil{Canadian Institute for Theoretical Astrophysics, University of Toronto, 60 St. George Street, Toronto, ON M5S 3H8, Canada}
\correspondingauthor{Wei Zhu}
\email{weizhu@cita.utoronto.ca}

\begin{abstract}
We study the influence of stellar metallicity on the fraction of stars with planets (i.e., the occurrence rate of planetary systems) and the average number of planets per star (i.e., the occurrence rate of planets). The former directly reveals the planet formation efficiency, whereas the latter reveals the final product of formation and evolution. We show that these two occurrence rates have different dependences on stellar metallicity. Specifically, the fraction of stars with planets rises gradually with metallicity, from $\sim$25\% to $\sim$36\% for 0.4 dex of [Fe/H] for all \emph{Kepler}-like planets (period $P<400$ days and radius $R_{\rm p}\gtrsim R_\oplus$). The average number of planets per star reaches a plateau (or possibly starts declining) at [Fe/H]$\gtrsim0.1$. This is plausibly caused by the emergence of distant giant planets at high metallicities, given that the close-in small planets and the distant giants preferentially co-exist in the same system. 
\end{abstract}

\keywords{methods: statistical --- planetary systems --- planets and satellites: general}

\section{Introduction} \label{sec:introduction}

Dependence of planet occurrence rate on host star properties provides insights into the formation of planet. The well established planet-metallicity correlation states that metal-rich stars preferentially host giant planets \citep{Santos:2001,Santos:2004,Fischer:2005}. This supports the core accretion model as the primary channel for giant planet formation (\citealt{IdaLin:2004}; but see \citealt{Nayakshin:2017} for an alternative explanation). However, there has been a long debate on whether or not such a correlation extends to smaller (radius $R_{\rm p}<4~R_\oplus$) planets \citep{Sousa:2008,Buchhave:2012,Buchhave:2014,Schlaufman:2015,Wang:2015,Buchhave:2015,Zhu:2016}. Putting aside the overall occurrence rate, several studies indeed have confirmed that small and hot (orbital period $P\lesssim10$ days) planets do appear preferentially around metal-rich stars \citep{Beauge:2013,Adibekyan:2013,Mulders:2016,Dong:2018,Petigura:2018,Wilson:2018}.

In studying the dependence on host star properties, one should be cautious about which occurrence rate to use. There are two types of occurrence rates to quantify the popularity of planets: the fraction of stars with planets, which we denote as $\Fp$, and the average number of planets per star, which we denote as $\barnp$. With the known number of stars, the former corresponds to the number of planetary systems and the latter the number of planets. The ratio, $\barnp/\Fp$, gives the average multiplicity, namely the average number of planets per planetary system \citep{Zhu:2018}. Therefore, unless every planetary system contains only one planet, the two occurrence rates would be different. Because the role of stellar properties comes primarily in determining the formation efficiency of planets, the fraction of stars with planets $\Fp$ is more appropriate to use \citep{Zhu:2016}. The average number of planets per star reveals the final product of both planet formation and dynamical evolution, the latter of which can largely modify the number of planets in the system. The problem becomes further complicated, because $\Fp$ is not a quantity that can be easily derived, in particular for transit missions such as \emph{Kepler} \citep{Youdin:2011,Zhu:2018}. For these reasons, previous statistical studies on the metallicity dependence all used the average number of planets per star either explicitly or implicitly \citep{Wang:2015,Buchhave:2015,Mulders:2016,Petigura:2018,Wilson:2018}. Therefore, their results do not necessarily reveal the dependence of the formation efficiency of small planets on stellar metallicity.

A recent work by \citet{Zhu:2018} combined the information of transiting planets and their non-transiting companions as inferred from the transit timing variations \citep{Holman:2005,Agol:2005} and constrained the fraction of Sun-like stars with \emph{Kepler}-like planets (periods $P<400$ days and radii $R_{\rm p}\gtrsim R_\oplus$) to be $\eta_{\rm Kepler}=30\pm3\%$. Using the methodology and the terminology of \citet{Zhu:2018}, we come to revisit the metallicity dependence of the small planet population.

This paper is organized as follows. In Section~\ref{sec:qualitative} we describe the data used in this work and compare the metallicity distribution of stars with and without planets. In Section~\ref{sec:quantitative} we derive the two occurrence rates ($\Fp$ and $\barnp$) for different metallicity bins. Our results are discussed in Section~\ref{sec:discussion}.

\begin{figure}[t]
\epsscale{1.2}
\plotone{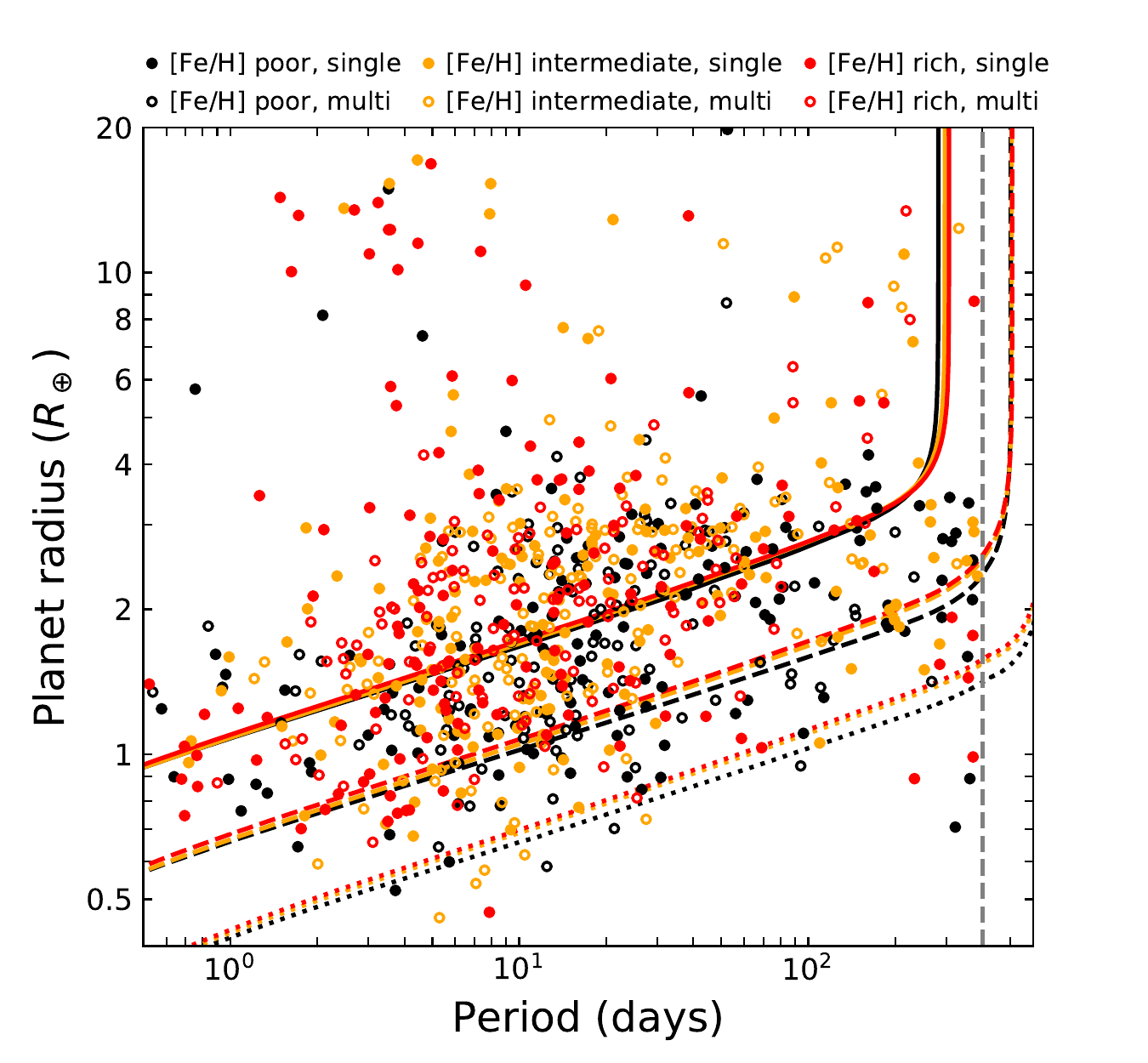}
\caption{Radii and orbital periods of planets found in the LAMOST-\emph{Kepler} sample. Planets are divided into six different categories depending on the host star metallicity and the number of transiting planets found. In particular, ``[Fe/H] poor'' means [Fe/H]$<-0.07$, ``[Fe/H] intermediate'' means $-0.07<$[Fe/H]$<0.10$, and ``[Fe/H] rich'' means [Fe/H]$>0.10$. These boundaries are chosen such that each bin contains the same number of planetary systems. We also over-plot the average detection efficiency curves for three metallicity bins, with solid, dashed, and dotted curves being the 90\%, 50\%, and 10\% efficiency curves, respectively.
\label{fig:sample}}
\end{figure}

\begin{figure}
\epsscale{1.1}
\plotone{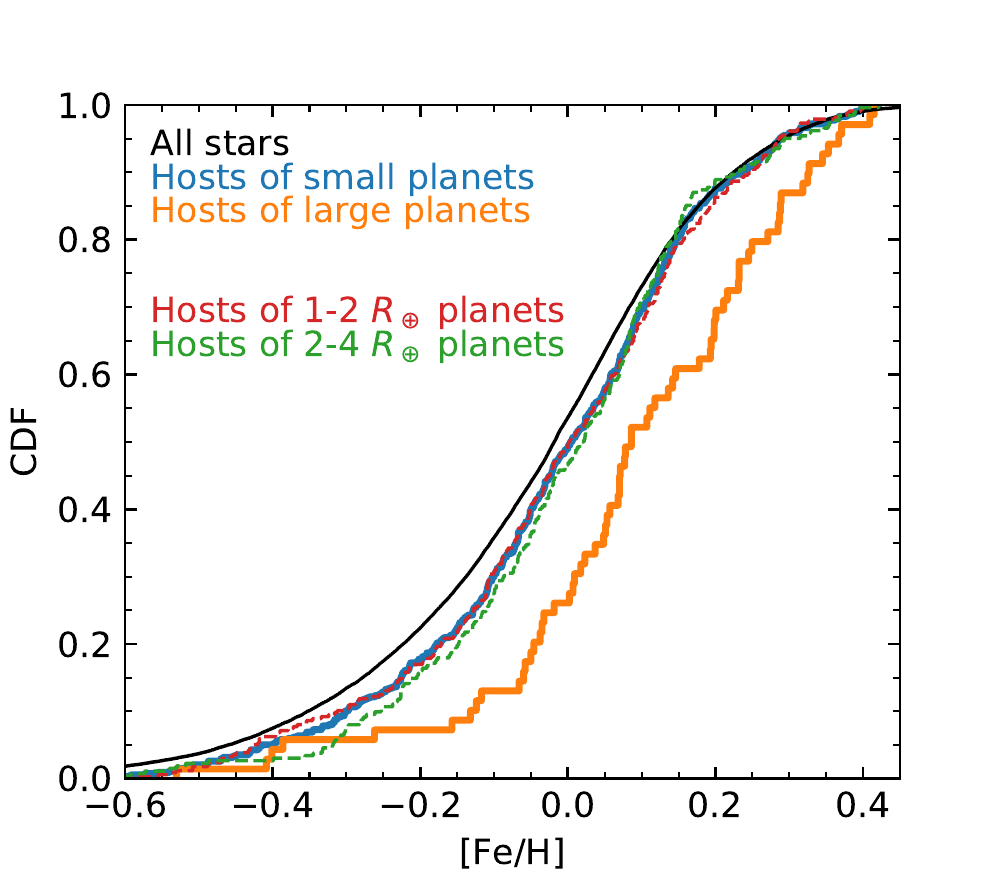}
\caption{Cumulative distribution functions (CDF) of metallicities of stars in different categories. Here small planets are planets with radii $R_{\rm p}<4~R_\oplus$, and large planets are those with $R_{\rm p}>4~R_\oplus$. Stars with planets belonging to multiple categories appear in all those categories, but stars with multiple planets in the same category are only counted once. This figure shows that planet hosts are preferentially metal-rich. There is no difference in metallicities of stars with 1-2 $R_\oplus$ planets and those with 2-4 $R_\oplus$ ones (sometimes called super-Earths and mini-Neptunes, respectively).
\label{fig:cdfs}}
\end{figure}

\section{\emph{Kepler} Planet Hosts Are Preferentially Metal-rich} \label{sec:qualitative}

A large and uniform spectroscopic survey of stars with and without planets is the key for the study of the metallicity dependence. We use the Sun-like stars sample from the \emph{Kepler}-LAMOST survey as refined in \citet{Zhu:2018}. The Large Sky Area Multi-Object Fiber Spectroscopic Telescope (LAMOST, also known as Goushoujing Telescope, \citealt{Cui:2012,Zhao:2012}), had surveyed and determined precise stellar parameters
\footnote{Here we use the parameters from the official LAMOST Stellar Parameter (LASP, \citealt{Luo:2015}) pipeline.}
for over 30\% of all \emph{Kepler} targets by 2017 (DR4), with no bias toward planet hosts \citep{Dong:2014,Decat:2015,Ren:2016,Xie:2016}. The refined Sun-like sample includes 827 transiting planets around 589 stars. We refer to \citet{Zhu:2018} for more details on the sample extraction.

Figure~\ref{fig:sample} shows the distribution of all planets in our sample. In this figure, planets are divided into different categories based on the host metallicity and the observed multiplicity. For demonstration purpose only, we choose three metallicity bins here, each with equal number of planetary systems. We also compute the averaged detection efficiency curves for individual metallicity bins using the code developed by \citet{Burke:2015}. As shown in Figure~\ref{fig:sample}, the detection efficiency curves are nearly indistinguishable, in particular for the two metal-rich bins. Similar conclusions were also found in \citet{Wang:2015} and \citet{Petigura:2018}. Therefore, from now on we will simply assume that the detection efficiency is the same for stars with a very broad range of bulk metallicities.

As a qualitative check, we first compare the cumulative distribution functions of metallicities of different types of stars, and the results are shown in Figure~\ref{fig:cdfs}. Hosts of large planets (here defined as planets with $R_{\rm p}>4~R_\oplus$) are systematically more metal-rich than hosts of small planets ($R_{\rm p}<4~R_\oplus$) and in particular the field stars, which again confirms the well-known giant planet-metallicity correlation \citep{Santos:2001,Santos:2004,Fischer:2005}. The hosts of small planets also appear to be systematically more metal-rich than field stars. A two-sample Kolmogorov-Smirnov (KS) test on these two metallicity samples gives $p=0.013$, indicating that the null hypothesis can be almost securely rejected that the two samples are drawn from the same underlying distribution. Furthermore, the metallicity distributions of stars with super-Earths ($1-2~R_\oplus$) and stars with mini-Neptunes ($2-4~R_\oplus$) are statistically the same (KS test $p=0.75$). This excludes any formation mechanism \citep[e.g.,][]{Dawson:2015} that would produce a distinction at $\sim2~R_\oplus$. It also suggests that the photo-evaporation mechanism, which produces the gap at approximately $2~R_\oplus$ \citep{Fulton:2017,OwenWu:2017,VanEylen:2018}, is probably not metallicity dependent (see also \citealt{Wu:2018}). Given that hosts of super-Earths and mini-Neptunes follow the same metallicity distribution, we only consider the distinction at $4~R_\oplus$ for the rest of the paper.

\section{Metallicity Dependence of Two Occurrence Rates} \label{sec:quantitative}

\begin{figure*}[b]
\epsscale{1.1}
\plotone{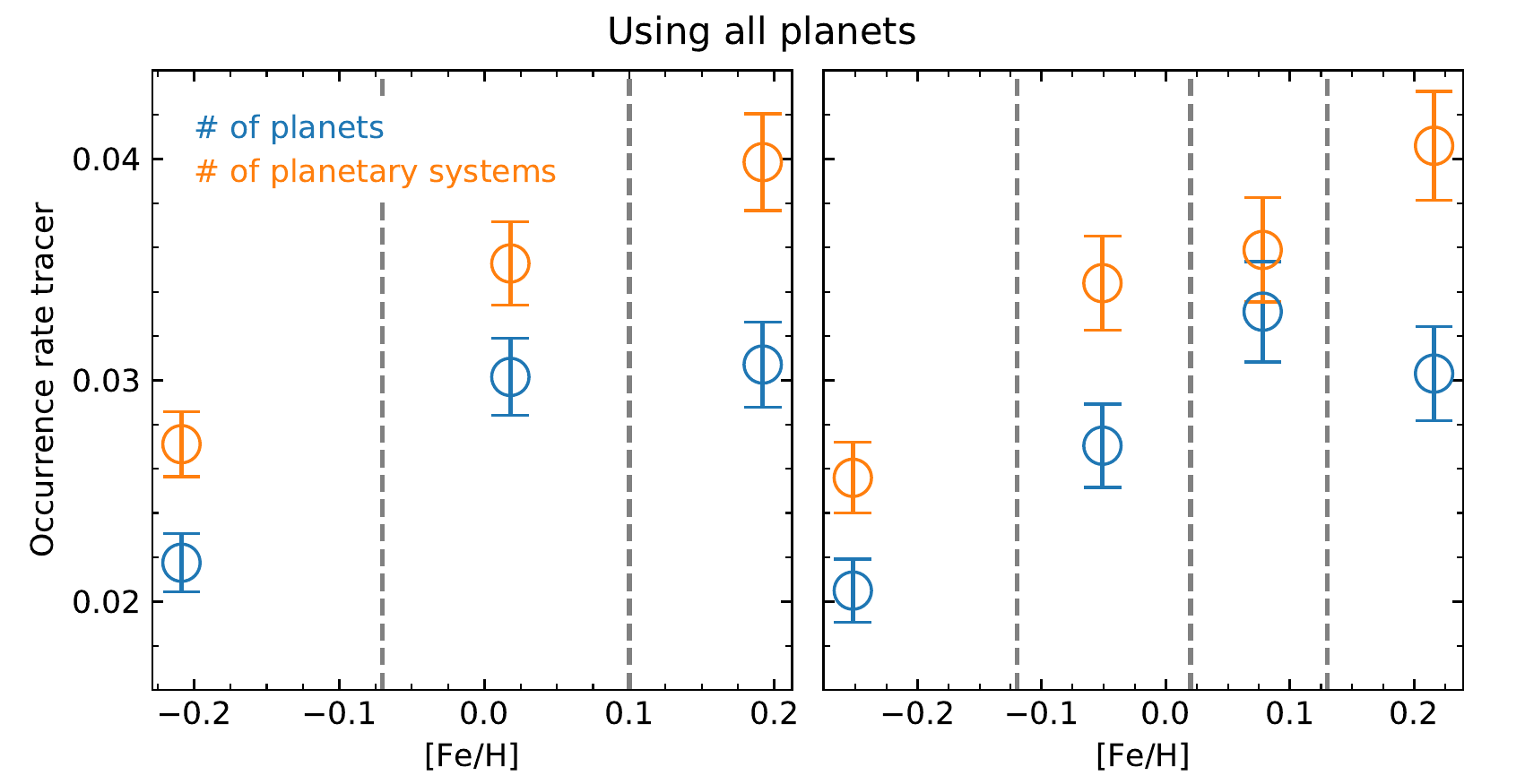}
\caption{The occurrence rate tracers (Equations~(\ref{eqn:np-tracer}) and (\ref{eqn:fp-tracer})) as functions of host star metallicities, determined from the \emph{Kepler}-LAMOST Sun-like sample. The whole sample is unevenly divided into three (left panel) or four (right panel) bins, such that each bin has nearly equal number of planetary systems. The vertical dashed lines indicate the dividing points of bins. In both panels, blue color is used for the occurrence rate of planets (i.e., the average number of planets per star $\barnp$), and orange color is used for the occurrence rate of planetary systems (i.e., the fraction of stars with planets $\Fp$). The $x$-axis positions are chosen as the median [Fe/H] values of planet hosts in individual bins. Here we have used all detected planets to compute the tracers.
\label{fig:tracers-1}}
\end{figure*}

Ideally, one wants to derive the two occurrence rates, the occurrence rate of planets ($\barnp$) and the occurrence rate of planetary systems ($\Fp$), as functions of stellar metallicity. However, even with the method of \citet{Zhu:2018} the current sample size is not large enough to constrain $\Fp$ reliably for multiple metallicity bins.

Alternatively, we use the following quantities as the \emph{tracers} of the two occurrence rates as derived by \citet{Zhu:2018}
\begin{equation} \label{eqn:np-tracer}
g_{11} \bar{n}_{\rm p} = \frac{1}{\mathcal{N}} \sum_{j=1}^K j N_j~;
\end{equation}
\begin{equation} \label{eqn:fp-tracer}
\langle g_{1k}+\sum_{j=1}^k g_{jk}\rangle F_{\rm p} = \frac{1}{\mathcal{N}} \left( N_1 + \sum_{j=1}^K N_j \right)\ .
\end{equation}
Here $g_{jk}$ is the probability that a $k$-planet system is seen to have $j$ transiting planets \citep{Tremaine:2012}, $\mathcal{N}$ is the total number of surveyed stars, $N_j$ is the observed number of systems with $j$ transiting planets, and $K$ is the maximum number of planets that one system can pack. The individual element $g_{jk}$ depends on not only the distribution of orbital periods (or more precisely, the transit parameters $\epsilon\equiv R_\star/a$) but also the distribution of orbital inclinations, except for $g_{11}$, which is given by
\begin{equation} \label{eqn:g11}
g_{11} = \langle \epsilon \rangle \equiv \int \epsilon f(\epsilon) \dif \ln{\epsilon}\ .
\end{equation}
Here $f(\epsilon)$ quantifies the (normalized) distribution of $\epsilon$ in logarithmic space. The combination $g_{1k}+\sum_{j=1}^k g_{jk}$ is largely insensitive to the choice of the input distributions of both inclinations and transit parameters. With the $\epsilon$ distribution derived from the current sample and the inclination distribution from \citet{Zhu:2018}, the typical values for the two prefactors are
\begin{equation}
g_{11} = 0.03;\quad \langle g_{1k}+\sum_{j=1}^k g_{jk}\rangle \approx 0.11\ .
\end{equation}

We divide the \emph{Kepler}-LAMOST sample into metallicity bins with uneven widths, each with the same number of planetary systems, and compute the $\barnp$ and $\Fp$ tracers for individual bins according to Equations~(\ref{eqn:np-tracer}) and (\ref{eqn:fp-tracer}). We derive the uncertainties by assuming that individual detections follow the Poisson distribution. We perform this analysis for different numbers of bins, to make sure that the trends we report below are not due to a specific choice of the number of bins.


Our results are shown in Figure~\ref{fig:tracers-1}. When taking the \emph{Kepler} planets as a whole, we find that the $\Fp$ tracer is consistently rising as [Fe/H] increases regardless of whether three or four metallicity bins are used. The $\barnp$ tracer seems to suggest a break in metallicity dependence: when metallicity is low, this tracer steadily increases with metallicity, but once the metallicity is relatively high, this tracer reaches a plateau (for 3-bin case) or even declines (for 4-bin case). Under the reasonable assumption that the actual occurrence rates follows these tracers, these trends suggest that the fraction of planetary systems consistently increases with stellar metallicities, whereas the average number of planets per star shows a break in the metallicity dependence.

\begin{figure*}[t]
\epsscale{1.1}
\plotone{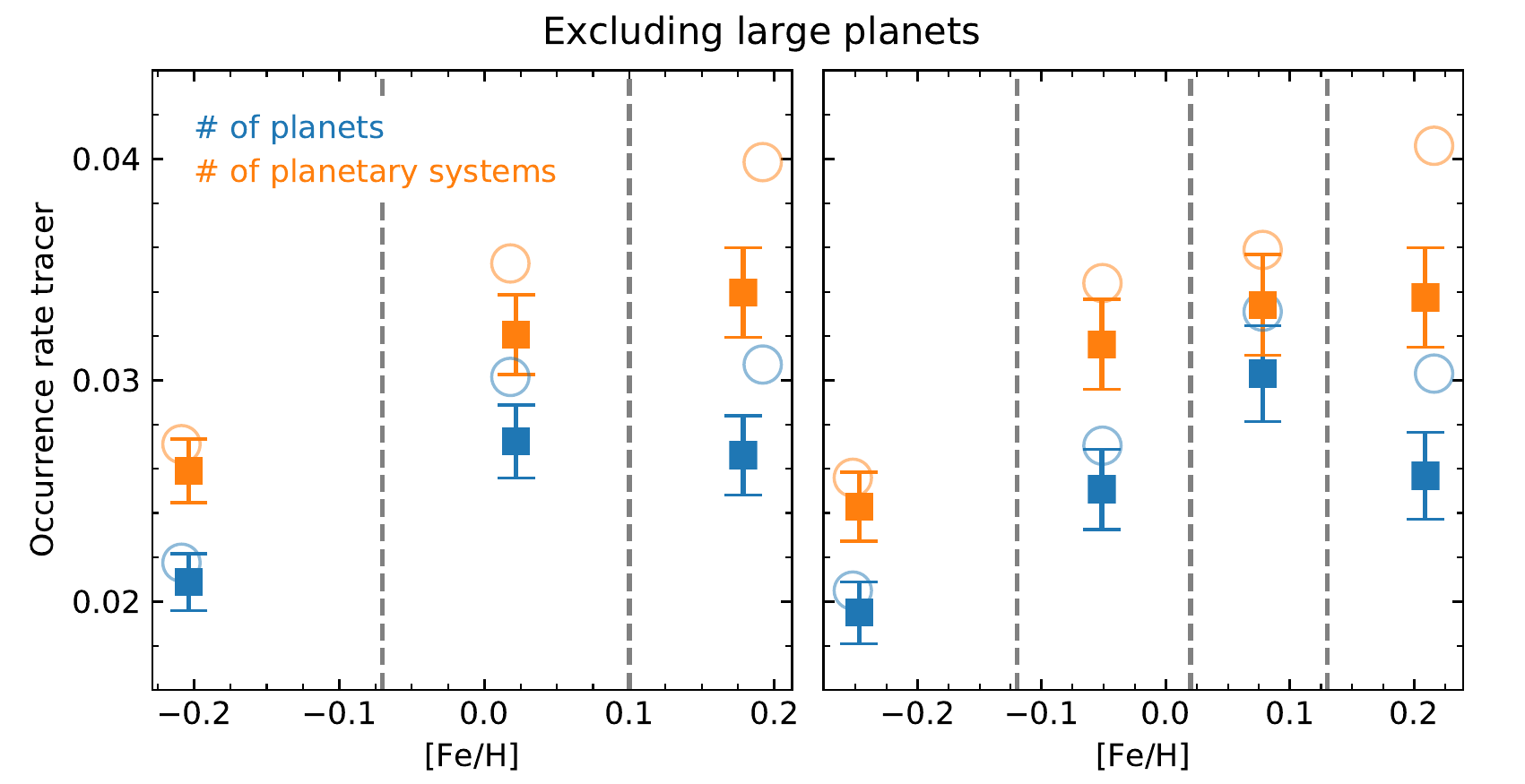}
\caption{The occurrence rate tracers as functions of stellar metallicity, after large planets (with $R_{\rm p}>4~R_\oplus$) are excluded. The open circles are the same as shown in Figure~\ref{fig:tracers-1}.
\label{fig:tracers-2}}
\end{figure*}

To be sure that the increasing behavior of the $\Fp$ tracer is not due to the known large planets, we also compute the two tracers after excluding planets with $R_{\rm p}>4~R_\oplus$. These include all Jupiter-sized planets and the majority of the hot Neptunes, which show strong dependence on stellar metallicity \citep{Dong:2018}. In principle, one can also choose to exclude systems with large planets, but because these large planets mostly reside in systems with single transiting planets (see Figure~\ref{fig:sample}), these two approaches do not make any significant difference. The results are shown as filled squares in Figure~\ref{fig:tracers-2}. As expected, the exclusion of the large planets modifies high-metallicity bins more than the low-metallicity ones, but the trends we see in the previous case remain: the $\Fp$ tracer is consistent with gradually increasing with metallicity, whereas the $\barnp$ tracer behaves differently with possibly a break at intermediate metallicities. We note that, if according to the right panel of Figure~\ref{fig:tracers-2}, the fraction of stars with small planets could be reaching a plateau since solar metallicity. This can potentially be explained as the saturation effect that is conceptualized in \citet{Zhu:2016}, but a larger data set is needed to confirm it.

\begin{figure*}
\epsscale{1.1}
\plotone{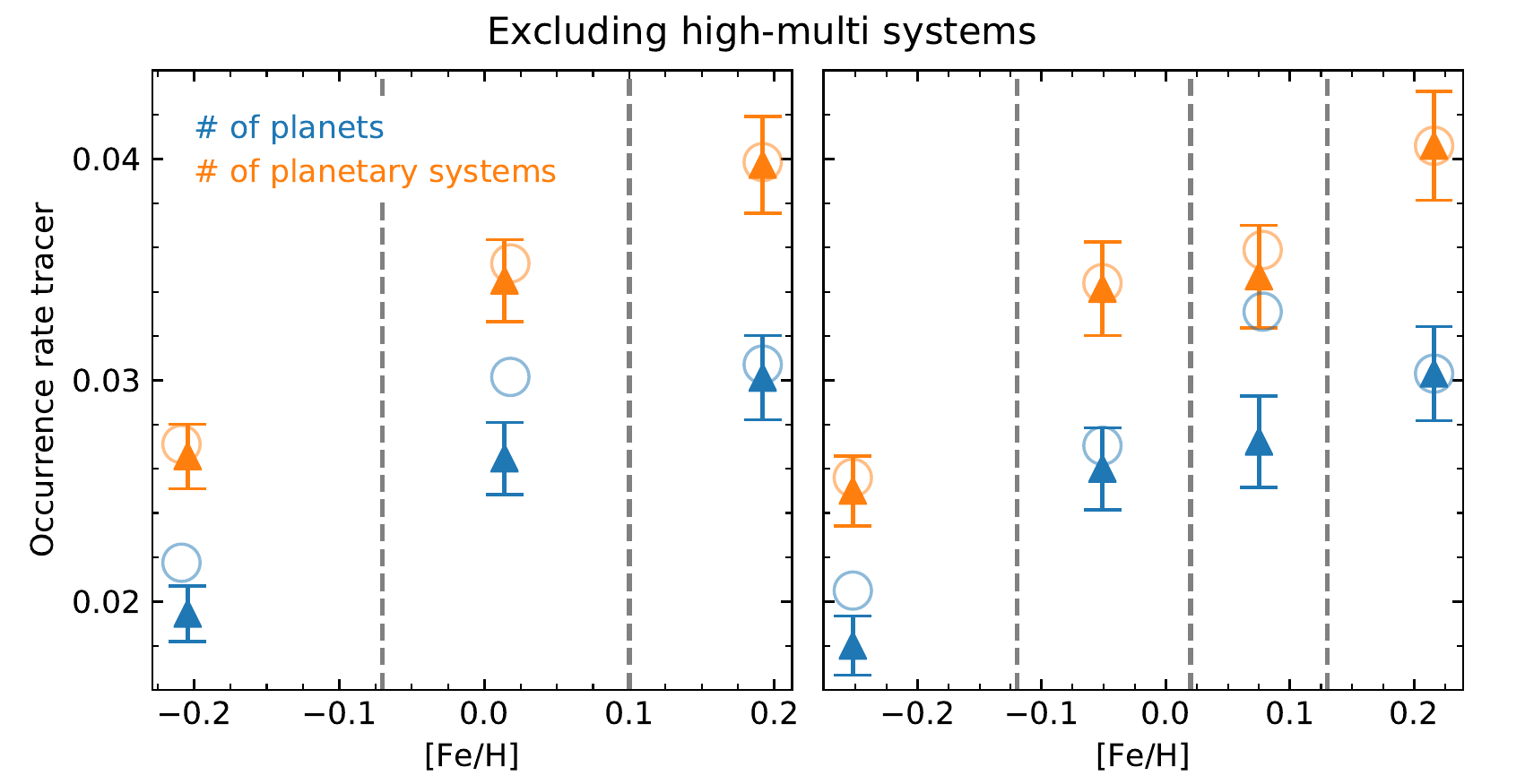}
\caption{The occurrence rate tracers as functions of stellar metallicity, after high-multiple systems (with at least four transiting planets) are excluded. The open circles are the same as shown in Figure~\ref{fig:tracers-1}. Large planets are included.
\label{fig:tracers-3}}
\end{figure*}

Finally, we study the influence of the high-multiple systems, which are systems with at least four transiting planets. Although there are only 16 (out of 589) high-multiple systems in our sample, they contribute in total 71 (out of 827) transiting planets. Therefore, these systems have larger weight on the $\barnp$ estimation than they do on the $\Fp$ estimation. We compute both tracers as functions of metallicities after excluding the high-multiple systems, and results are shown as filled triangles in Figure~\ref{fig:tracers-3}. As expected, the $\Fp$ tracer is largely unaffected by the exclusion of high-multiple systems, but the break that we saw in the metallicity dependence of the $\barnp$ tracer in previous cases disappears. Specifically, the high-multiple systems concentrate in the metallicity range $0\lesssim$[Fe/H]$\lesssim0.1$ (see Figure~4 of \citealt{ZhuWu:2018} for an illustration), and thus the removal of them affects the $\barnp$ tracer in this intermediate metallicity range the most. Such a concentration appears as a consequence of dynamical evolution rather than formation efficiency: there may have been more high-multiple systems at higher metallicities, but these systems were largely destroyed because of the emergence of cold giants. See Section~\ref{sec:discussion} for further explanations. After the high-multiple systems are excluded, the fraction of systems with planets and the average number of planets per star follow similar trends, and both are monotonically increasing with stellar metallicity.

\section{Discussion} \label{sec:discussion}

The occurrence rate of giant planets shows strong dependence on stellar metallicity, but previous studies have not agreed on whether this correlation extends to small planets. In this work, we revisit this issue with the help of large and uniform spectroscopic LAMOST survey of the \emph{Kepler} stars. We point out that previous studies have commonly used the occurrence rate of planets, measured as the average number of planets per star $\barnp$, when studying the metallicity dependence (and mass dependence as well). However, this occurrence rate reflects the final outcome of planet formation and evolution, and is therefore not a direct measure of the formation efficiency. The occurrence rate of planetary system, measured as the fraction of stars with planets $\Fp$, reflects more directly the planet formation efficiency.

To study the metallicity dependence, one therefore would like to derive $\Fp$ as a function of stellar metallicity. However, because of the difficulty in constraining $\Fp$ precisely with the available data, we use the quantity $\langle g_{1k}+\sum_{j=1}^k g_{jk} \rangle \Fp$ as a \emph{tracer} of $\Fp$. For comparison, we also study the $\barnp$ tracer $g_{11}\barnp$. We derive these two tracers for different stellar metallicities, and find that they do behave differently in particular at high stellar metallicities. Specifically, the $\Fp$ tracer continuously increases as stellar metallicity increases, whereas the $\barnp$ tracer shows a break in its metallicity dependence. We show that these trends hold even if only small planets are included.

It is plausible that the correlation between close-in small planets and distant giant planets drives the difference in the two occurrence rates. As \citet{ZhuWu:2018} recently found,
\footnote{See also \citet{Bryan:2018} and \citet{Herman:2018}.}
about one third of the \emph{Kepler}-like systems have distant cold Jupiters, and this fraction doubles in the high-metallicity ([Fe/H]$>0.1$) systems. The emergence of cold Jupiters at high metallicities can drive dynamical instabilities in systems that are already heavily packed, which reduces the number of planets but not the number of planetary systems \citep[e.g.,][]{Matsumura:2013,Huang:2017,Lai:2017,Pu:2018}. After the exclusion of systems with at least four transiting planets, the break in the metallicity dependence of the $\barnp$ tracer does disappear. The correlation between inner and outer planetary systems can also explain more detailed trends in the metallicity dependence of the occurrence rate $\barnp$. \citet{Petigura:2018} found that the occurrence rate of large planets consistently increases with stellar metallicity, but the occurrence rate of relatively small planets shows different behavior: within orbital period $P\sim10$ days, the occurrence rate consistently increases; beyond that the occurrence rate seems to decrease with metallicity. This, in principle, could also be explained by the correlation between inner small planets and the cold giants, as smaller planets are more likely affected by the cold giants \citep{Lai:2017}.

Our work indicates that the formation efficiency of relatively small planets still depends on the stellar metallicity, but this dependence is much weaker than the giant planet-metallicity correlation. Taking the nominal value $\langle g_{1k}+\sum_{j=1}^k g_{jk} \rangle\approx 0.11$, we find that $\Fp$ changes from $25\%$ at [Fe/H]$\approx-0.2$  to $36\%$ at [Fe/H]$\approx+0.2$ and thus by only a factor of 1.4, whereas the giant planet occurrence changes by a factor of $\sim10^{2\Delta{\rm [Fe/H]}}=6.3$. This is also consistent with the qualitative behaviors of the cumulative distributions of metallicities as shown in Figure~\ref{fig:cdfs}. Therefore, although the formation efficiencies of both small and giant planets depend on stellar metallicities, giant planets require more stringent conditions for their formation.

\acknowledgements
I would like to thank Yanqin Wu and Subo Dong for discussions as well as comments on the manuscript. I also thank the anonymous referee for comments.
W.Z. was supported by the Beatrice and Vincent Tremaine Fellowship at CITA.
This paper includes data collected by the \emph{Kepler} mission. Funding for the \emph{Kepler} mission is provided by the NASA Science Mission directorate.
This paper also uses data from the LAMOST survey. Guoshoujing Telescope (the Large Sky Area Multi-Object Fiber Spectroscopic Telescope, LAMOST) is a National Major Scientific Project built by the Chinese Academy of Sciences. Funding for the project has been provided by the National Development and Reform Commission. LAMOST is operated and managed by the National Astronomical Observatories, Chinese Academy of Sciences.


\end{CJK*}

\begin{thebibliography}{}
\bibitem[Adibekyan et al.(2013)]{Adibekyan:2013} Adibekyan, V.~Z., Figueira, P., Santos, N.~C., et al.\ 2013, \aap, 560, A51 
\bibitem[Agol et al.(2005)]{Agol:2005} Agol, E., Steffen, J., Sari, R., \& Clarkson, W.\ 2005, \mnras, 359, 567 
\bibitem[Beaug{\'e} \& Nesvorn{\'y}(2013)]{Beauge:2013} Beaug{\'e}, C., \& Nesvorn{\'y}, D.\ 2013, \apj, 763, 12 
\bibitem[Bryan et al.(2018)]{Bryan:2018} Bryan, M.~L., Knutson, H.~A., Fulton, B., et al.\ 2018, arXiv:1806.08799
\bibitem[Buchhave et al.(2014)]{Buchhave:2014} Buchhave, L.~A., Bizzarro, M., Latham, D.~W., et al.\ 2014, \nat, 509, 593 
\bibitem[Buchhave \& Latham(2015)]{Buchhave:2015} Buchhave, L.~A., \& Latham, D.~W.\ 2015, \apj, 808, 187 
\bibitem[Buchhave et al.(2012)]{Buchhave:2012} Buchhave, L.~A., Latham, D.~W., Johansen, A., et al.\ 2012, \nat, 486, 375 
\bibitem[Burke et al.(2015)]{Burke:2015} Burke, C.~J., Christiansen, J.~L., Mullally, F., et al.\ 2015, \apj, 809, 8 
\bibitem[Cui et al.(2012)]{Cui:2012} Cui, X.-Q., Zhao, Y.-H., Chu, Y.-Q., et al.\ 2012, Research in Astronomy and Astrophysics, 12, 1197 
\bibitem[Dawson et al.(2015)]{Dawson:2015} Dawson, R.~I., Chiang, E., \& Lee, E.~J.\ 2015, \mnras, 453, 1471 
\bibitem[De Cat et al.(2015)]{Decat:2015} De Cat, P., Fu, J.~N., Ren, A.~B., et al.\ 2015, \apjs, 220, 19 
\bibitem[Dong et al.(2018)]{Dong:2018} Dong, S., Xie, J.-W., Zhou, J.-L., Zheng, Z., \& Luo, A.\ 2018, Proceedings of the National Academy of Science, 115, 266
\bibitem[Dong et al.(2014)]{Dong:2014} Dong, S., Zheng, Z., Zhu, Z., et al.\ 2014, \apjl, 789, L3 
\bibitem[Fischer \& Valenti(2005)]{Fischer:2005} Fischer, D.~A., \& Valenti, J.\ 2005, \apj, 622, 1102 
\bibitem[Fulton et al.(2017)]{Fulton:2017} Fulton, B.~J., Petigura, E.~A., Howard, A.~W., et al.\ 2017, \aj, 154, 109 
\bibitem[Herman et al.(2019)]{Herman:2018} Herman, M.~K., Zhu, W., \& Wu, Y.\ 2019, arXiv:1901.01974
\bibitem[Holman \& Murray(2005)]{Holman:2005} Holman, M.~J., \& Murray, N.~W.\ 2005, Science, 307, 1288 
\bibitem[Huang et al.(2017)]{Huang:2017} Huang, C.~X., Petrovich, C., \& Deibert, E.\ 2017, \aj, 153, 210 
\bibitem[Ida \& Lin(2004)]{IdaLin:2004} Ida, S., \& Lin, D.~N.~C.\ 2004, \apj, 616, 567 
\bibitem[Lai \& Pu(2017)]{Lai:2017} Lai, D., \& Pu, B.\ 2017, \aj, 153, 42 
\bibitem[Luo et al.(2015)]{Luo:2015} Luo, A.-L., Zhao, Y.-H., Zhao, G., et al.\ 2015, Research in Astronomy and Astrophysics, 15, 1095 
\bibitem[Matsumura et al.(2013)]{Matsumura:2013} Matsumura, S., Ida, S., \& Nagasawa, M.\ 2013, \apj, 767, 129 
\bibitem[Mulders et al.(2016)]{Mulders:2016} Mulders, G.~D., Pascucci, I., Apai, D., Frasca, A., \& Molenda-{\.Z}akowicz, J.\ 2016, \aj, 152, 187 
\bibitem[Nayakshin(2017)]{Nayakshin:2017} Nayakshin, S.\ 2017, \pasa, 34, e002 
\bibitem[Owen \& Wu(2017)]{OwenWu:2017} Owen, J.~E., \& Wu, Y.\ 2017, \apj, 847, 29 
\bibitem[Petigura et al.(2018)]{Petigura:2018} Petigura, E.~A., Marcy, G.~W., Winn, J.~N., et al.\ 2018, \aj, 155, 89 
\bibitem[Pu \& Lai(2018)]{Pu:2018} Pu, B., \& Lai, D.\ 2018, \mnras, 478, 197 
\bibitem[Ren et al.(2016)]{Ren:2016} Ren, A., Fu, J., De Cat, P., et al.\ 2016, \apjs, 225, 28 
\bibitem[Santos et al.(2001)]{Santos:2001} Santos, N.~C., Israelian, G., \& Mayor, M.\ 2001, \aap, 373, 1019 
\bibitem[Santos et al.(2004)]{Santos:2004} Santos, N.~C., Israelian, G., \& Mayor, M.\ 2004, \aap, 415, 1153 
\bibitem[Schlaufman(2015)]{Schlaufman:2015} Schlaufman, K.~C.\ 2015, \apjl, 799, L26 
\bibitem[Sousa et al.(2008)]{Sousa:2008} Sousa, S.~G., Santos, N.~C., Mayor, M., et al.\ 2008, \aap, 487, 373 
\bibitem[Tremaine \& Dong(2012)]{Tremaine:2012} Tremaine, S., \& Dong, S.\ 2012, \aj, 143, 94 
\bibitem[Van Eylen et al.(2018)]{VanEylen:2018} Van Eylen, V., Agentoft, C., Lundkvist, M.~S., et al.\ 2018, \mnras, 479, 4786 
\bibitem[Wang \& Fischer(2015)]{Wang:2015} Wang, J., \& Fischer, D.~A.\ 2015, \aj, 149, 14 
\bibitem[Wilson et al.(2018)]{Wilson:2018} Wilson, R.~F., Teske, J., Majewski, S.~R., et al.\ 2018, \aj, 155, 68 
\bibitem[Wu(2018)]{Wu:2018} Wu, Y.\ 2018, arXiv:1806.04693 
\bibitem[Xie et al.(2016)]{Xie:2016} Xie, J.-W., Dong, S., Zhu, Z., et al.\ 2016, Proceedings of the National Academy of Science, 113, 11431 
\bibitem[Youdin(2011)]{Youdin:2011} Youdin, A.~N.\ 2011, \apj, 742, 38 
\bibitem[Zhao et al.(2012)]{Zhao:2012} Zhao, G., Zhao, Y.-H., Chu, Y.-Q., Jing, Y.-P., \& Deng, L.-C.\ 2012, Research in Astronomy and Astrophysics, 12, 723 
\bibitem[Zhu \& Wu(2018)]{ZhuWu:2018} Zhu, W., \& Wu, Y.\ 2018, \aj, 156, 92 
\bibitem[Zhu et al.(2016)]{Zhu:2016} Zhu, W., Wang, J., \& Huang, C.\ 2016, \apj, 832, 196 
\bibitem[Zhu et al.(2018)]{Zhu:2018} Zhu, W., Petrovich, C., Wu, Y., Dong, S., \& Xie, J.\ 2018, \apj, 860, 101 
\end{thebibliography}
\end{document}